\documentclass[twocolumn,showpacs,preprintnumbers,amsmath,amssymb]{revtex4}

\usepackage{graphicx}
\usepackage{dcolumn}
\usepackage{bm}
 
\begin{document}

\title{Density functional theory for freezing transition of vortex-line liquid
with periodic layer pinning}

\author{Xiao Hu\(^{*}\), Mengbo Luo\(^{*,\dagger}\), and Yuqiang Ma\(^{*,\ddagger}\)}
\affiliation{ \(^{*}\)Computational Materials Science Center,
National Institute for Materials Science, Tsukuba 305-0047, Japan \\
\(^{\dagger}\)Department of Physics, Zhejiang University, Hangzhou 310027, 
People's Republic of China \\
\(^{\ddagger}\)Department of Physics, Nanjing University, Nanjing 210093, 
People's Republic of China
}

\date{\today}

\begin{abstract}
By the density functional theory for crystallization, it is shown that for vortex lines
in an underlying layered structure a smectic phase with period \(m=2\) can be stabilized 
by strong layer pinning.  The freezing of vortex liquid is then 
two-step, a second-order liquid-smectic transition and a first-order smectic-lattice 
transition.  DFT also indicates that a direct, first-order liquid-lattice transition
preempts the smectic order with \(m\ge 3\) irrespectively of 
the pinning strength.  Possible \(H-T\) phase diagrams are mapped out.  Implications of 
the DFT results to the interlayer Josephson vortex system in high-\(T_c\) cuprates are given.

\end{abstract}


\maketitle


 It is now well established that the Abrikosov vortex lattice in type II superconductors
\cite{Abrikosov} melts via a thermodynamic first-order transition \cite{reviews}.  
However, much less consensus has been
reached on the possible phases and melting process of interlayer Josephson vortex lattice
where layer pinning is essential.  A Lorentz-force independent dissipation was found in 
Bi\(_2\)Sr\(_2\)CaCu\(_2\)O\(_{8+y}\) at \(H=5T\) parallel to the \(ab\)-plane \cite{Iye}, 
which is possibly a signature of a Kostlitz-Thouless (KT) phase 
\cite{Efetov,Ivlev,Blatter2}.  On the other hand, transport experiments conducted in 
YBa\(_2\)Cu\(_3\)O\(_{7-\delta}\) suggested a second-order softening of the vortex lattice \cite{Kwok},
which can be accounted for by a smectic phase \cite{Balents}.  
A recent computer simulation found continuous 
meltings and a KT-type intermediate phase above a multicritical field \cite{HuPRL,HuPRB}.  

As a general theory for crystallization, the density functional approach was formulated by
Ramakrishnan and Yussouff \cite{RY} (See also \cite{Ramakrishnan}).  
The basic idea of the density functional theory (DFT) is to describe the lattice
phase and the freezing transition by the direct pair correlation function (DPCF) in
liquid phase, which in turn can be evaluated from macroscopic quantities such as the average
density and temperature.  At the clean limit, both the first-order nature of the vortex-line 
liquid freezing and quantitative aspects such as the Lindemann
number have been successfully captured by DFT \cite{Sengupta}.
Later on, DFT was used to investigate vortex-line systems with point-like \cite{Menon} 
and columnar \cite{Dasgupta} pinning centers.  

In the present work, we apply DFT to explore the freezing of vortex-line liquid
subject to periodic layer pinning \cite{Tachiki,Barone}.
As it is hard to evaluate the DPCF with finite 
interlayer Josephson coupling \cite{Sengupta}, which is essential
in the present case, we will not try to draw the detailed phase diagram. Instead, 
we adopt the DPCF and the layer pinning strength as parameters to explore the possible
freezing processes.  This approach turns out to be fruitful:
A vortex smectic phase with more vortex lines in every other layers (period \(m=2\)) is 
observed at strong layer 
pinning.  When the smecitc phase is present, the freezing of vortex liquid is two-step, a 
second-order liquid-smectic and a first-order smectic-lattice transition. 
A direct, first-order liquid-lattice transition preempts the smectic order with \(m\ge 3\)
irrespectively of the pinning strength.   Possible topologies of \(H-T\)
phase diagrams are figured out.  


The free energy of a vortex-line system measured from the uniform liquid is given by 
\cite{Ramakrishnan,Sengupta,Chakrabarti}

\begin{equation}
 \beta \Delta {\cal F}=
\int d^2r\sum_n \left[ \rho_n({\bf r})\ln\frac{\rho_n({\bf r})}{\rho_0}
                      +(A-1) \delta\rho_n({\bf r}) \right] \nonumber
\end{equation}

\vspace{-5mm}

\begin{equation}
\hspace{13mm}
-\frac12\int d^2rd^2r'\sum_{n,n'} \delta\rho_n({\bf r})\delta\rho_{n'}({\bf r}')  
                                       C({\bf r}-{\bf r}',n-n') \nonumber
\end{equation}

\vspace{-5mm}

\begin{equation}
\hspace{-15mm} -\int d^2r\sum_{n} \delta\rho_n({\bf r}) \beta V_{\rm p} \cos(qz),
\end{equation}

\noindent where \(\delta \rho({\bf r})= \rho({\bf r}) - \rho_0\) with \(\rho_0\) 
the average areal vortex density and \(C({\bf r}-{\bf r}',n-n') \) the DPCF.  
The summations are taken over vortex segments along the 
magnetic field (\(B\| {\bf y}\)), and the integrals over the transverse directions
{\bf x} and {\bf z}, with \({\bf z} \| c\) axis.  \(V_{\rm p}\) is the pinning energy per vortex
segment.

By means of the variational calculus, the free-energy density per vortex segment of a vortex 
lattice is found as

\begin{equation}
 \frac{\beta \Delta {\cal F}}{N_{xz}N_y}
= -A + \frac12 s_{\rm u} \sum_{\bf K} \rho_{\bf K}^2 C({\bf K},0),
\end{equation}

\noindent with \(s_{\rm u}=1/\rho_0\) the area of real-space unit cell (u.c.), \(N_{xz}\)
the number of unit cells and \(N_y\) the number of vortex segments,
the summation over the reciprocal lattice vectors (RLVs), 
\(\rho({\bf r})=\rho_0+\sum_{\bf K} \rho_{\bf K} \exp(i{\bf K}\cdot{\bf r})\), and
\( C({\bf k},0) = \int d^2r \sum_n  C({\bf r},n) \exp(-i{\bf k}\cdot{\bf r})\).
The chemical potential \(A\) is given by

\begin{equation}
A=\ln \frac1{s_{\rm u}}\int_{\rm u.c.} d^2r e^{ 
                     \sum_{\bf K} \rho_{\bf K} C({\bf K},0) \exp(i{\bf K}\cdot{\bf r})
                    +\beta V_{\rm p} \cos(qz) },
\end{equation}

\noindent with the integral over the unit cell.
The lattice phase is characterized by the non-vanishing Fourier components of the 
vortex density, which thus serve as the order parameters for the freezing transition.
Following previous studies \cite{Ramakrishnan,Sengupta}, we adopt a few order parameters
in the present work confirming that more will not change the conclusions qualitatively.

Because of the linear coupling with the vortex density, the layer
pinning \(V_{\rm p}\) appears only in the chemical potential \(A\) in the
free-energy density.  
The layer pinning is effective when the pinning wavenumber \(q\) coincides with
one of the RLVs.  As the DPCF \(C({\bf K},0)\) decreases
quickly as going to higher shells of RLVs, the layer pinning is most 
important when its wave vector coincides with one of the primitive RLVs.

For weak magnetic fields, vortices reside in a periodic subset of block layers with
period \(ms\) where \(s\) is the layer spacing. 
For large \(m\), the pinning energy term in Eq.(3) 
decouples from the lattice-order term and only modifies the
measure of the integral of the lattice potential.  In such a 
case the layer pinning is irrelevant to the phase transition.  The freezing process
is then essentially the same with the Abrikosov lattice with the scaling by the anisotropy
parameter \cite{Blatter3}.  It takes numerics shown below to see to what extension that 
the large-\(m\) limit governs the freezing phenomenon .  

\vspace{2mm}
\noindent {\it Vortex segment and layer pinning energy--}
When a single vortex line is placed in a periodic layer pinning potential, a kink (kernel) presumes 
a metastable state in which a part of the vortex line cross the pinning barrier and reside
in the neighbor energy valley.  The crossing
takes place within the length scale \(L_{\rm wall}\simeq s\sqrt{\gamma\epsilon_0 /4U_{\rm p}}\),
where \(\gamma=\lambda_{\rm c}/\lambda_{\rm ab}\), 
\(\epsilon_0=(\phi_0/4\pi\lambda_{\rm ab})^2\), \(U_{\rm p}\) the pinning energy per
unit length \cite{Barone,Balents}. The pinning energy is  
\(V_{\rm p}\simeq s \sqrt{\gamma \epsilon_0 U_{\rm p}}\), and the kink length defined as 
the separation between two crossings is 
\(L_{\rm kink}=L_{\rm wall}\exp(2\beta V_{\rm p})\) \cite{Balents}.  
The physical choice of the vortex segment in the free-energy functional (1) is therefore the kink
with the pinning energy per kink \(V_{\rm p}\).  The above estimates are physically meaningful  
when \(L_{\rm kink}\) is sufficiently larger than \(L_{\rm wall}\), or equivalently
\(\beta V_{\rm p}\) is large enough.    Numerically, for YBa\(_2\)Cu\(_3\)O\(_{7-\delta}\)
with \(T_c=92K\), \(\kappa=100\), \(\gamma=8\), \(\lambda_{ab}(0)=1000\AA\) and \(s=12\AA\), one has 
\(L_{\rm wall}\simeq 10s\), \(L_{\rm kink}\simeq 60s\), and \(\beta V_{\rm p}\simeq 0.8\) at
\(T=88K\).  As temperature decreases, \(\beta V_{\rm p} \) increases linearly to 
\(\beta V_{\rm p}\simeq 4.3\) at \(T=80K\), \(L_{\rm wall}\) decreases slowly and
\(L_{\rm kink}\) increases exponentially.  For Bi\(_2\)Sr\(_2\)CaCu\(_2\)O\(_{8+y}\)
with \(T_c=90K\), \(\kappa=100\), \(\gamma=150\), \(\lambda_{ab}(0)=2000\AA\) and \(s=15\AA\),  
one has \(\beta V_{\rm p}\simeq 0.2\) at \(T=84K\), which increases linearly to 
\(\beta V_{\rm p}\simeq 0.3\) at \(T=78K\). 

\begin{figure}
\vspace{3cm}
\includegraphics{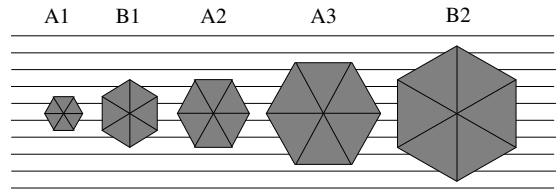}
\caption{\label{fig:unitcell}
Real-space unit cells commensurate with layer modulation.  
The \(x\) direction is rescaled with the anisotropy parameter \(\gamma\).
A and B are for the type of unit cell, and the number denotes \(m\) defined in text.  
}
\end{figure}

\vspace{2mm}
\noindent {\it Lattice unit cell--}
The vortex lattice structure in the presence of layer pinning is nontrivial.  In an anisotropic
material {\it without} layer modulation, there are two stable directions of triangle vortex 
lattice, which are described by the following two unit cells \cite{Ivlev}:  
Type A: \({\bf a}_1=C\sqrt{\gamma} {\bf x}, 
{\bf a}_2=\frac{C\sqrt{\gamma}}{2} {\bf x} + \frac{C\sqrt{3} }{2\sqrt{\gamma} } {\bf z}\);
Type B: \({\bf a}_1=\frac{C}{\sqrt{\gamma}} {\bf z}, 
{\bf a}_2=\frac{C\sqrt{3\gamma} }{2} {\bf x} +  \frac{C}{2\sqrt{\gamma}} {\bf z}\),
with \(C^2=2\phi_0/\sqrt{3}B\).  The associated RLVs are
Type A: \({\bf b}_1=\frac{2\pi}{C\sqrt{\gamma}} {\bf x}
-\frac{2\pi \sqrt{\gamma}} {C\sqrt{3}} {\bf z}, 
{\bf b}_2=\frac{4\pi\sqrt{\gamma}}{C\sqrt{3}} {\bf z}\);
Type B: \({\bf b}_1=-\frac{2\pi}{C \sqrt{3\gamma}} {\bf x} 
 +\frac{2\pi\sqrt{\gamma}  }{C} {\bf z},
  {\bf b}_2=\frac{4\pi}{C\sqrt{3\gamma}} {\bf x}\). 

In DFT the reduction of free energy of crystallization is evaluated by the DPCF
of liquid, which is isotropic after the space-rescaling with the anisotropy 
parameter.  Just above the freezing, there appears a sharp peak in the DPCF at 
\(k_0=4\pi/\sqrt{3}C\) associated with the uniform vortex density of the rescaled, 
{\it isotropic} liquid.  The reduction of free energy is maximal when 
the RLVs of a candidate lattice coincide with \(k_0\) in magnitude.  It is easy to 
see that the two lattice structures listed above match this condition.

On the other hand, the vortex lattice is requested to be commensurate with the underlying 
layer modulation in DFT, which is reasonable for layered superconductors like
high-\(T_c\) cuprates.  This makes the pinning energy equivalent for different lattice
structures.  Therefore, at the following sequence of the magnetic fields
\(H_1=\sqrt{3}\phi_0/2\gamma s^2\), \(H_2=\phi_0/2\sqrt{3} \gamma s^2\), 
\(H_3=\sqrt{3}\phi_0/8\gamma s^2\), \(H_4=\phi_0/6\sqrt{3} \gamma s^2\), 
and \(H_5=\phi_0/8\sqrt{3} \gamma s^2\), ..., the vortex lattices depicted in Fig. 1 
resume minimal free energy.  
The stable lattice structure at magnetic fields nearby those listed above
should be the same with those in Fig. 1 except for that the vortex separation in the \(x\) 
direction is adjusted in order to accommodate the vortex density.  
The ground-state lattice structure changes when two freezing DPCFs coincide.  
The lattice structure with a smaller freezing DPCF is achieved at a higher temperature 
and is stable for the given magnetic field.
Possible first-order transition between different lattices upon sweeping the magnetic field
at fixed temperature was addressed in Ref.\cite{Bulaevskii}.

\vspace{2mm}
\noindent {\it Freezing transitions--}
We now proceed to investigate the freezing process and nature of phase transition when the
lattice structure is given and the DPCF is swept.  This corresponds to reducing temperature at given
magnetic fields.  We will focus on the sequence of magnetic fields listed above, where
the vortex lattice are {\it naturally} commensurate with the layer structure.  Although analyses on other
magnetic fields are {\it technically} cumbersome as \(C({\bf K},0)\)'s assume different values even
on the same shells of RLVs, the possible phases and nature of phase transition should be the same.
As read from Eqs. (2) and (3), freezing transitions for different anisotropy parameter \(\gamma\)
are governed by the same set of \(C({\bf K},0)\)'s as functions of \(V_{\rm p} \) and \(m\).  This
corresponds to the scaling property addressed in Ref.\cite{Blatter3},
apart from the layer pinning effect.

For A1 in Fig. 1, the present system is equivalent to colloids under laser radiation 
addressed in Ref.\cite{Chakrabarti}.  As the pinning energy increases,
the freezing switches from first order to continuous at a tricritical point at
\(\beta V_{\rm p} \simeq 0.212\) and \(\rho_0 C^{(1)}\simeq 0.748\). The notation with \(C^{(i)}\) referring to
\(C({\bf K},0)\) for \({\bf K}\) on the \(i\)-th shell of RLVs is adopted in the present Letter.

\begin{figure}
\vspace{5.5cm}
\includegraphics{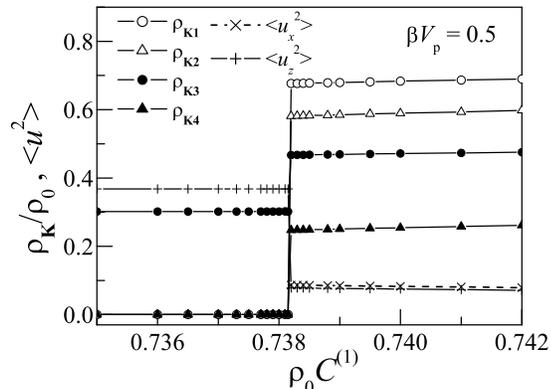}
\caption{\label{fig:A2-1}
First-order freezing into the lattice structure A2 for \(\beta V_{\rm p} =0.5\), with  
\(C^{(3)}=0.3C^{(1)}\).  Thermal fluctuations \(\langle u^2_x \rangle\) and 
\(\langle u^2_z \rangle\) are normalized by \((\gamma s)^2\) and \(s^2\) respectively.
\(\langle u^2_x \rangle=\infty\) for \(\rho_0 C^{(1)}<0.7382\).
}
\end{figure}

\begin{figure}
\vspace{5.5cm}
\includegraphics{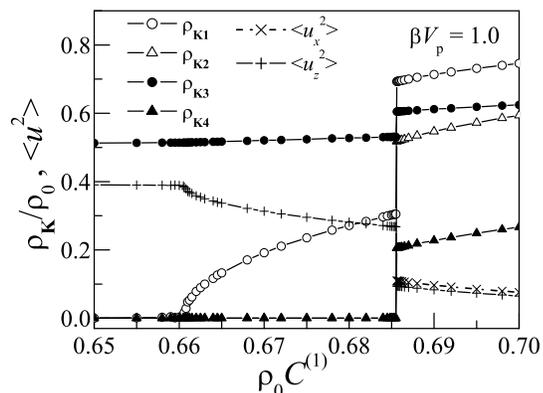}
\caption{\label{fig:A2-2}  
Two-step freezing, a second-order and a first-order transitions, into the lattice 
structure A2 and an intermediate smectic phase for \(\beta V_{\rm p} =1\), with \(C^{(3)}=0.3C^{(1)}\).  
Thermal fluctuations \(\langle u^2_x \rangle\) and 
\(\langle u^2_z \rangle\) are normalized by \((\gamma s)^2\) and \(s^2\) respectively.
\(\langle u^2_x \rangle=\infty\) for \(\rho_0 C^{(1)}<0.6856\).
}
\end{figure}

For A2, we use four order parameters \(\rho_{\bf K}\) where 
\( {\bf K}1= {\bf b}_2\), \({\bf K}2={\bf b}_1\), \({\bf K}3= 2{\bf b}_2\), and 
\({\bf K}4= 2{\bf b}_1\) with \({\bf b}_2=\pi/s {\bf z}\).  
A finite \(\rho_{{\bf K}3}\) characterizes the modulated liquid.  The    
freezing is first order at low pinning energy such as \(\beta V_{\rm p} =0.5\) depicted in Fig. 2, 
where all order parameters except for the trivial one set up discontinuously. 
At large pinning energy, a phase characterized by 
\(\rho_{{\bf K}1}>0\) and \(\rho_{{\bf K}2}=\rho_{{\bf K}4}=0\) appears 
via a second-order phase transition as shown explicitly in Fig. 3 for \(\beta V_{\rm p} =1\). 
The vortex density modulation is associated with a wave number half 
of the underlying layer structure, corresponding to more vortices in every
other layers; the system behaves as liquid in in-plane directions.  
This is the smectic order addressed in Ref.\cite{Balents}, and the liquid-smectic 
transition is in the Ising universality class since \(m=2\). 
As DPCF increases, the full lattice order appears via a first-order transition.
The \(\rho_0C^{(1)}-\beta V_{\rm p} \) phase diagram for A2 is presented in Fig. 4, with a multicritical point
at \(\rho_0C^{(1)}\simeq 0.714\) and \(\beta V_{\rm p} \simeq 0.712\).  This is one of the central
observations of the present DFT study.

\begin{figure}
\vspace{5.5cm}
\includegraphics{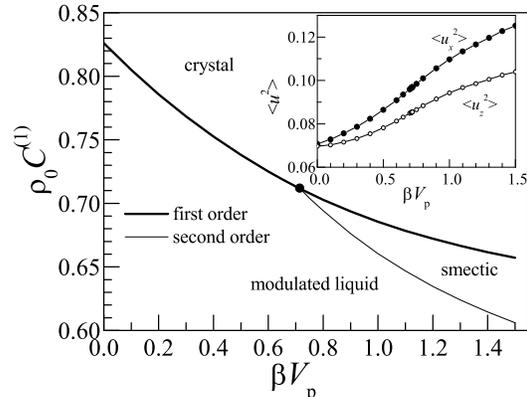}
\caption{\label{fig:pdA2}
\(\rho_0C^{(1)}-\beta V_{\rm p}\) phase diagram for the lattice A2.  Inset: mean squared
displacements, normalized by \((\gamma s)^2\) and \(s^2\) in the \(x\) and
\(z\) directions respectively, evaluated along the first-order phase boundary. 
}
\end{figure}

Similar analysis has been performed for A-type lattice with \(m\ge 3\), using 
\( C^{(2)} =0.5 C^{(1)}\), \( C^{(3)} =0.3 C^{(1)}\), \( C^{(5)} =0.1 C^{(1)}\) and so on.  
In contrast with A1 and A2, we cannot find
the smectic phase and continuous transition up to huge pinning energy 
accessible in numerics. This DFT prediction is different from the analysis in Ref.\cite{Balents}.

For unit cell of type B, the crystallization is always one step and first order.  In this case no 
primitive Bragg peak matches with the layer modulation, and thus the layer 
pinning is much ineffective compared with the lattices A1 and A2.  As the result, the six 
primitive Bragg peaks appear simultaneously which is associated with a first-order transition. 
The first-order melting transition observed in Monte Carlo simulations for \(B=\phi_0/32s^2\) 
and \(\gamma=8\) \cite{HuPRL,HuPRB} is in accordance with the DFT, since the lattice structure
is B1. In principle, there might be a smectic phase associated with the two Bragg spots 
\({\bf K}=(0,\pm \pi/s)\) on the second shell of RLVs for B2.  However, numerical 
analysis reveals that this situation is not realized since the layer pinning is at the
6th shell of RLVs and the interaction with those on the 2nd shell is very weak.

\begin{figure}
\vspace{7cm}
\includegraphics{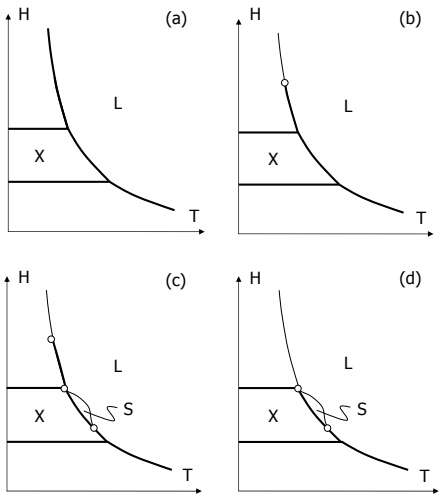}
\caption{\label{fig:HTpd}
Possible \(H-T\) phase diagrams for vortex lines in periodic layer pinning potential from
DFT, with L, X and S abbreviating liquid, lattice and smectic respectively. 
The thick (thin) curves denote first(second)-order phase boundaries.  Multicritical
points are marked by open circles.  The horizontal phase boundaries are drown only for
B1-A2 (A1-A2 in (d)) and A2-A3 transitions. 
}
\end{figure}

\vspace{2mm}
\noindent {\it Phase diagram--}
Although the detailed freezing curve cannot be drawn without calculations on DPCF at given 
magnetic fields and temperatures, we are able to categorize possible \(H-T\) phase 
diagrams based on the above analyses. Figure 5(a) corresponds to a system of weak layer 
pinning, in which freezing is always first order.  The two 
horizontal phase boundaries are between lattices structures B1-A2 and A2-A3 \cite{Bulaevskii}. 
Note that similar phase boundaries may be observed at lower magnetic fields, and that
the lattices A1 and B1 can be tuned into each other through crossover.
In Fig. 5(b), the freezing transition into vortex lattice A1 is of second order
at high magnetic fields where the freezing temperature is low and
the layer pinning is strong enough.  At even stronger layer pinning, a smectic phase 
appears and transforms into lattice A2 upon cooling as shown in Fig. 5(c).  As the magnetic
field increases to the regime where the lattice B1 presumes the ground state, the smectic phase 
shrinks to zero.  It is possible practically that the lattice B1 
is unstable comparing with other lattice structures, resulting in Fig. 5(d).  
No further complex phase diagram is expected from DFT.

\vspace{2mm}
\noindent {\it Discussions--}
Based on the DFT analysis and experimental phase diagrams 
for \(H<10T\) \cite{Kwok,Schilling}, the \(m=2\) smectic phase is expected for 
YBa\(_2\)Cu\(_3\)O\(_{7-\delta}\) around \(H=H_3\simeq 39T\).  No smectic phase is plausible
for \(H\le H_4=\phi_0/6\sqrt{3}\gamma s^2\simeq 17T\).  Layer pinning may be too weak
in Bi\(_2\)Sr\(_2\)CaCu\(_2\)O\(_{8+y}\) to stabilize the smectic.

Since DFT only treats the RLVs, it is unable to describe quasi long-range orders.
A KT phase is therefore out of the scope where DFT can reach.  With this
drawback of DFT in mind, one may notice that the tricritical point for A1 lattice in Fig. 5,
which should exist in both YBa\(_2\)Cu\(_3\)O\(_{7-\delta}\) and 
Bi\(_2\)Sr\(_2\)CaCu\(_2\)O\(_{8+y}\), 
corresponds to the multicritical point above which an intermediate KT phase is 
observed in computer simulations \cite{HuPRB}.  

As displayed in the inset of Fig. 4, thermal fluctuations are more anisotropic than
the anisotropy parameter \(\gamma\), which is clearly caused by the layer pinning.
This anisotropy exists irrespectively of 
the existence of smectic phase, and thus cannot be taken as a precursor
of a partial melting \cite{XHComment}.  In presence of
layer pinning the Lindemann numbers are not constant any more.

One may expect that the results derived above
apply to two-dimensional (2D) systems \cite{Ramakrishnan} with periodic line pinning. 
It is noticed however that thermal fluctuations are more important in 2D as addressed
in Ref.\cite{Frey}.

\vspace{2mm}
\noindent {\it Acknowledgements--}
One of us (X.H.) thanks Masashi Tachiki for useful discussions.
This work was partially supported by Japan Society for the
Promotion of Science (Grant-in-Aid for Scientific Research (C) No. 15540355).

\end{document}